\documentclass[prb,noshowpacs,superscriptaddress,twocolumn]{revtex4-1}

\usepackage{amsmath}
\usepackage{graphicx}
\usepackage{bm}

\usepackage{latexsym}

\usepackage{amsmath}
\usepackage{amsfonts}
\usepackage{graphicx}
\usepackage{multirow}
\usepackage{chngcntr}

\usepackage{color}
\usepackage[usenames,dvipsnames]{xcolor}
\usepackage{natbib}
\usepackage[final=true]{hyperref}
\hypersetup{
    colorlinks=true,
    linkcolor=Maroon,
    filecolor=magenta,      
    urlcolor=cyan,
    citecolor=PineGreen,    
}

\newcommand{\rnd}{\mathrm{rnd}^{(0..1)}}
\newcommand{\myf}{1$^{\mathrm{st}}$}
\newcommand{\mys}{2$^{\mathrm{nd}}$}

\begin{document}
\title{Effects of resource competition on evolution and adaptive radiation}
\author{S.~V.~Koniakhin}
\email{kon@ibs.re.kr}
\affiliation{Center for Theoretical Physics of Complex Systems, Institute for Basic Science (IBS), Daejeon 34126, Republic of Korea}
\affiliation{Basic Science Program, Korea University of Science and Technology (UST), Daejeon 34113, Korea}

\date{\today}
%
\begin{abstract}
The entanglement of population dynamics, evolution, and adaptive radiation for species competing for resources is studied. For resource harvesting, we modify the model used in Ref. \textit{Phys. Rev. Lett.} \textbf{118} 048103 and introduce new resource contest principles. We realistically implement the effects of beneficial and deleterious mutations on the coefficients in the equations governing the population dynamics and consider the emergence of reproductive isolation. The proposed model is shown to be in agreement with the competitive exclusion principle and no vacant niche principle. We establish a mechanism that contributes to preventing the accumulation of irreversible deleterious mutations: competition between recently diverged species/subpopulations. The proposed model is applicable for descriptions of more complex systems. In case of many constants in time resources, one observes very rapid specialization, a feature not reproducible by the common model.

\end{abstract}

\maketitle


\section{Introduction}


The processes of speciation and adaptive radiation (AR) play crucial roles in evolution and the appearance of diversity of life forms on Earth. AR exists at various levels of taxonomy and periods of natural history. It starts from the last unified common ancestor (LUCA) that originated the fundamental current domains Archaea, Bacteria, and Eukarya. Another example of AR is the Cambrian explosion when all the principal body plans of currently living animals were founded. Another noticeable example is the AR in mammals after the Cretaceous–Paleogene extinction event. Finally, the most comprehensively studied example of evolution and speciation is the evolution of the \textit{Homo} genus and related forms since approximately 3 million years to the present day. One essential aspect of evolution and speciation is the competition between ecologically close species or even between subpopulations recently subjected to reproductive isolation. This process can only be described by taking into account both population dynamics (with a crucial role of resource competition) and fitness dynamics due to mutations and natural selection pressure. To support the recent outstanding progress in our knowledge of evolution from the so-called fossil record, genome and morphological analyses of living organisms, and paleogenomics, appropriate mathematical tools and models are highly desired.

Basic models for population dynamics and interspecies competition include the simplest in the sense of mathematical formulation logistic equations~\cite{PhysRevE.69.021908,PhysRevE.94.042413,law2003population}, predator--prey models~\cite{PhysRevE.88.062721,PhysRevE.95.042404,PhysRevE.98.022410,akcakaya1995ratio}, and replicator equations~\cite{PhysRevE.70.061914,PhysRevLett.89.148101}. Numerics is typically used to solve these models, but in some cases analytical approaches have been applied~\cite{PhysRevE.69.021908,PhysRevE.98.022410,bagnoli1997speciation,PhysRevE.94.022423}. Sometimes more complicated systems of equations for species abundance and resource/prey growth rates are used to study population dynamics in more detail~\cite{rosenzweig1963graphical,mac1969species,revilla2009multispecies,grover1997resource}. Such models can also be extended to account for spacial distributions~\cite{PhysRevE.94.042413,banerjee2017spatio,law2003population}. In the most comprehensive case, whole food webs~\cite{PhysRevE.88.062721,PhysRevE.97.022404,roopnarine2010networks} for real ecosystems (scaling from a Petri dish to lakes and oceans~\cite{martinez1991artifacts}) have been investigated. Theoretical studies of food webs allow quantifying global phenomena related to macro evolution like extinction avalanches~\cite{PhysRevLett.82.652}. The advanced Webworld model~\cite{caldarelli1998modelling,drossel2001influence} can successfully reproduce such parameters of real food webs as the number of trophic levels, amount of links per species, and fractions of basal/intermediate/top species. Investigation into evolution is also accessible using some abstract analytical models~\cite{PhysRevE.98.012405,bagnoli1997speciation} and numerical individual-based models (IBMs)~\cite{PhysRevE.63.031903,PhysRevE.69.051912,markov2012can,PhysRevE.74.021910,chow2004adaptive,brigatti2007evolution}. Population traits can be traced via replicator equations~\cite{PhysRevLett.105.178101,PhysRevE.84.051921}. Among other evolutionary phenomena, sexual selection~\cite{PhysRevE.74.021910,van2009origin,yoshimura1994population} and the problem of cooperation~\cite{PhysRevLett.95.238701,PhysRevLett.105.178101,PhysRevE.84.051921} have been of interest. One can employ advanced genetic simulators~\cite{sanford2012next,hoban2012computer,peng2005simupop} like SimuPOP or Mendels accountant to provide maximal approximation to real genetic mechanisms~\cite{foll2014widespread}. However, focusing on the full genome in genetic simulators can be impractical due computational complexity and complicated results interpretation.

The great variety of additional models and significant extensions of existing ones has been proposed to account for various aspects of population dynamics and evolutionary mechanisms. For example, it was shown that AR and speciation can be simulated~\cite{PhysRevE.72.031916} using the approach of diffusing gas in trait morphospace with the probability of extinction positively correlated with the local density of species. In Refs.~\cite{PhysRevE.63.031903,PhysRevE.69.051912} applying IBM approaches on a finite lattice, it was demonstrated that the combination of stochastic inheritance and natural selection for fitness gives significant advantage to population protection against extinction with respect to cases without evolution. Previously, a related model was used to demonstrate the negative effect of eugenics on genetic diversity resulting in extinction after drastic changes in conditions~\cite{cebrat1999model,pkekalski2000effect} as well as to study migrations to new habitats~\cite{mroz1996conditions,mroz1999model,hui2018invade}. Ref.~\cite{brigatti2005evolution} accounted for the genomes of individuals and studied speciation in the case of competition depending on distance in phenotype space. An explicit number of resource types limiting abundance and thus originating competition was the essential feature of the model employed in Refs. \cite{PhysRevLett.118.048103,10.7554/eLife.15747}. The obtained result was the existence of a phase transition leading to a collective state decoupled from external conditions.


 Most current research tends to study either ecological aspects (population dynamics and ecosystem stability) or evolutionary aspects. However, a number of interesting investigations have combined population dynamics with evolution, allowing the study of their mutual effects~\cite{PhysRevE.98.012405,PhysRevLett.95.078105,billiard2016stochastic,PhysRevLett.105.178101,PhysRevE.84.051921}. The above-mentioned food web studies using the Webworld model~\cite{caldarelli1998modelling,drossel2001influence} also include evolution aspects. These models can be simplified, though, for the case of ecologically close species recently subjected to reproductive isolation with a weak role of predator--prey relationships. The aim of this paper is to develop and explore a model for the simultaneous description of population dynamics, speciation, and their mutual influence that correctly accounts for the following factors: 1) competition for resources affecting the population dynamics, 2) beneficial and deleterious mutations leading to fluctuations in resource harvesting effectiveness, and 3) appearance of reproductive isolation between populations.

The paper is organized as follows. In Section~\ref{sec_model} the model is introduced. The first part of the model description is devoted to the population dynamics and resource distribution model (subsection~\ref{subs_res}), the second part accounts for the particular implementation of beneficial and deleterious mutations (subsection~\ref{subs_mutations}), and the third part describes the speciation process (subsection~\ref{subs_Spec}). In the Results, Section~\ref{sec_res}, we demonstrate that the present model correctly reproduces the following important ecological concepts. The first is the competitive exclusion principle, postulating that each ecological niche is filled with only one species (subsection~\ref{subs_competexcl}). Then we demonstrate a resistance against the accumulation of deleterious mutations: overall fitness increases even when the rate of deleterious mutations overcomes that of beneficial mutations (subsection~\ref{subs_decay}). Further, we demonstrate that the model is in agreement with the absence of vacant ecological niches principle (subsection~\ref{subs_emptyniche}). Finally, the model was employed to describe AR in complex systems evolving in the presence of a large number of accessible resource types in cases of stable and unstable influx (subsection~\ref{subs_specialization}). In Section~\ref{sec_concl} the conclusion is given.




\section{Formulating the model}
\label{sec_model}

\subsection{Resource competition and population dynamics}
\label{subs_res}

As a basic model for resource competition we start from the one used by Tikhonov and Monasson in Ref.~\cite{PhysRevLett.118.048103}, which is closely related to the classic model from Ref.~\cite{mac1969species}. In this model, all resources positively contribute to the maximum achievable species abundance; Liebig's law of the minimum describes a different situation. It is natural to employ the notations from Ref.~\cite{PhysRevLett.118.048103}:

- $n_{\mu}$ is species abundance (population) $\mu$ at simulation step $s$

- $M$ is the total number of species at step $s$

- $\sigma_{i\mu}$ is the harvesting effectiveness of the resource of type $i$ by species $\mu$. In Ref. \cite{PhysRevLett.118.048103} this quantity is referred to as ``investment of species $\mu$ into harvesting resource $i$''

- $R_i$ is the total influx of the resource of type $i$

- $N$ is the number of resource types

- $h_{\mu i}$ is the availability of the resource of type $i$ for species $\mu$

For the population dynamics, we begin with the conventional equation, 
\begin{equation}
    \frac{dn_{\mu}}{dt} = n_{\mu} f(\Delta_{\mu}).
\end{equation}
In practice at each step $s$ we do the following,
\begin{equation}
\label{eq_kinteic2}
    n_{\mu}^{(s+1)} = n_{\mu}^{(s)}(1 + f(\Delta_{\mu})),
\end{equation}
where $\Delta_{\mu}$ is the resource surplus for species $\mu$, and $f(x)$ gives the rate of population growth as a function of resource surplus. As far as we study dynamical systems not in equilibrium, the particular form of $f(x)$ is important to obtain the actual species abundance growth rate. In contrast to Ref.~\cite{PhysRevLett.118.048103}, the equilibrium state $\frac{dn_{\mu}}{dt}=0$ is studied and thus only the conditions of $f(0)=0$ and monotonic growth are required. The shape of $f(x)$ used here satisfies these conditions, and its particular form is discussed below.

The resource surplus $\Delta_{\mu}$ is defined by the amount of total resource influx, the species harvesting effectiveness, and the species abundance as

\begin{equation}
    \Delta_{\mu} = \sum_i\sigma_{\mu i}h_{\mu i} - \chi_{\mu},
    \label{eq_surplus}
\end{equation}
where $h_{\mu i}$ is availability of resource $i$ for species $\mu$, and $\chi_{\mu} = 1$ is the requirement of the resource by species $\mu$.

In Ref.~\cite{PhysRevLett.118.048103} the availability of each resource is common for all species ($h_{\mu i}\equiv h_i$) and depends only on the total resource influx and its total demand by all species. The total species abundance in the ecosystem is defined by the total resource influx (in fact the total carrying capacity) but not by the typical amplitude of the species harvesting effectiveness. For the case of a single species and single resource within the framework of Ref.~\cite{PhysRevLett.118.048103} (all indices can be omitted due to unambiguity), the availability is written $h=R/(n\sigma)$, resource surplus is $\Delta = R/n - 1$, and maximum abundance is $n_{\textrm{max}}=R$ regardless of the actual fitness (harvesting effectiveness) of the species.

In contrast, we require the model of resource competition to demonstrate an increase in the total abundance in the ecosystem when the fitness (harvesting effectiveness) increases. This can be implemented by setting $R_i\sigma_{\mu i}$ as an influx of resource $i$ accessible by species $\mu$. This requirement leads to the following formulation of the model~\cite{misccodes}.

\begin{figure}
\includegraphics[width=0.45\textwidth]{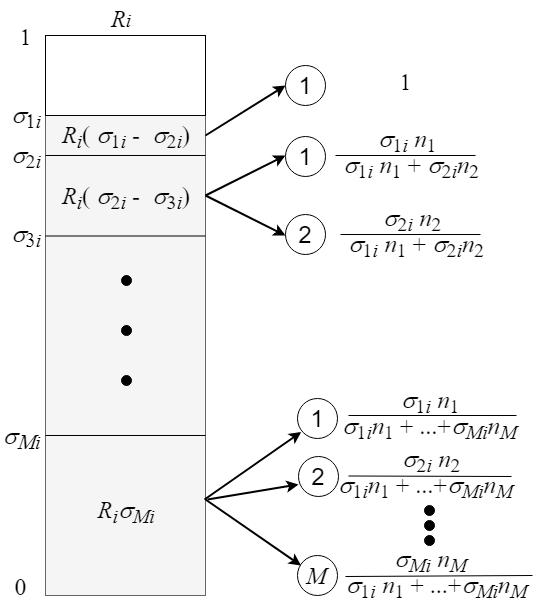}
\caption{\label{fig1_distribution} The employed principles of resource competition and distribution. For harvesting each resource $i$, the species are sorted by their harvesting effectiveness: $\sigma_{1i}>\sigma_{2i}>...>\sigma_{Mi}$. The total influx of the $i$-th resource $R_i$ is denoted as the vertical rectangle, which is separated into smaller rectangles symbolizing the resource fractions accessible by various species. The upper gray rectangle corresponds to the part of the resource accessible by the fittest species in harvesting this resource, followed by the parts of resource accessible together by the the \myf~and the \mys~species, etc. The lower rectangle reflects the resource part accessible by all species up to $M$ (least fit). The labels within the rectangles show the corresponding fractions of resource influx, and the labels of the circles show the fractions of sharing the corresponding rectangle among the species (their sum always equals 1 for each rectangle). This scheme is to be independently applied for all resources $i=1..N$.}
\end{figure}

To calculate resource availability $h_{\mu i}$ for resource $i$, we first sort the species by descending $i$-th resource harvesting effectiveness: $\sigma_{1i}>\sigma_{2i}>...>\sigma_{Mi}$. After sorting, the species with the largest harvesting effectiveness has no competitor for the part of resource $i$ that is the most difficult to gain, and hence harvests $R_i\cdot(\sigma_{1i}-\sigma_{2i})$ solely. Then the species begin competition for the remaining resource $i$. The \myf~and the \mys~species share the next part of the $i$-th resource, namely $(\sigma_{2i} - \sigma_{3i})R_i$. Each species income is proportional to its harvesting effectiveness and abundance, which matches Ref.~\cite{PhysRevLett.118.048103}. Hence, the \myf~species harvests $R_i\sigma_{1i}n_1/(\sigma_{1i}n_1+\sigma_{2i}n_2)$ and the \mys~species harvests $R_i\sigma_{2i}n_2/(\sigma_{1i}n_1+\sigma_{2i}n_2)$. The next species are then included, and the next accessible parts of the resource are distributed. At the last step, $R_i\cdot \sigma_{Mi}$ is shared among all species. This principle is illustrated in Fig.~\ref{fig1_distribution}. Such sorting of harvesting effectiveness $\sigma_{\mu i}$ and consequential resource distribution should be performed for all resources independently. In terms of the resource availability $h_{\mu i}$, the proposed model can be written as follows:

\begin{equation}
h_{\mu i} = R_i \sum_{\mu'=\mu}^{M} (\sigma_{\mu'i}-\sigma_{\mu'+1i}) S_{\mu' i}^{-1},
\label{eq_hmui}
\end{equation}
where
\begin{equation}
   S_{\mu' i} = \sum_{\mu''=1}^{\mu'}\sigma_{\mu''i}n_{\mu''}.
\end{equation}

To preserve mathematical correctness, one should introduce the auxiliary variable $\sigma_{\mu,M+1}=0$.

In the case of a single species and single resource, one will obtain $h=R/n$ and a surplus of $\Delta = \sigma R/n-1$, and thus the limit value of abundance is $\sigma R$. While the principles of resource distribution between species employed in the present model are less trivial than in Ref.~\cite{PhysRevLett.118.048103}, one essential feature of the present model is that it requires an increase in harvesting effectiveness for increasing the accessible parts of the resource and consequentially increasing species abundance. Moreover, the model also provides stronger competition between species, making specialization more favorable.

We use the following form of function $f(x)$,
\begin{equation}
    f(x)=
    \begin{cases}
x, & x\leq0\\
x/2,& 0<x<0.4\\
0.2,& x\geq0.4
\end{cases}.
\end{equation}
First, this obeys the monotonic growth and $f(0)=0$ conditions. The function shape at negative $x$ restricts the maximum abundance at the level of carrying capacity in the situation of resource shortage. Taking the slope 0.5 for $0<x<0.4$ mitigates the population growth process as a response to a high resource influx. The maximum value of the function is saturated, which restricts the maximum abundance growth rate at the level of 20\% per step, coinciding with the time discretization used in Ref.~\cite{drossel2001influence}. In practice, the unitary step can be considered here as one generation or several generations; e.g., there is an estimation of 20\% per 1000 years of population growth for early \textit{Homo}~\cite{khrisanfova2005anthropology}. The equations are formulated as dimensionless, but quantities with naturally dimensionalized units can be used as well~\cite{misc1}.

Together with Eq.~\eqref{eq_kinteic2}, the developed resource competition model behaves within the framework of the ratio-dependent functional response~\cite{arditi1989coupling,akcakaya1995ratio,abrams2000nature,drossel2001influence,berryman1992Origins}: a constant speed of abundance growth at low consumer population, and further saturation of consumer abundance at a level proportional to resource influx. The resources do not experience negative feedback of the species intake like in Ref.~\cite{mac1969species} where the concept of the limited renewal rate was used. Here, the resource influx is either constant in time or regulated according to the simulation design.

\subsection{Mutations}
\label{subs_mutations}

This section describes the implementation of mutation events and consequential harvesting effectiveness fluctuations essential for the evolution process. Typically, due to natural selection the near-optimum position in a trait morphospace is reached and maintained in a population, and it is much simpler to corrupt existing well-functioning mechanisms than to invent something new and well working~\cite{behe2004simulating}. Thus, fitness-affecting mutations can be divided into two classes. One corresponds to small deleterious mutations, from which the fitness of the species slowly decreases. The second mutation class is beneficial mutations, which are rarer and lead to an increase in fitness.

Both types of mutations are modeled as fluctuations in harvesting effectiveness~\cite{PhysRevE.69.021908}. At each step, we set $\sigma_{\mu i}^{(s+1)} =\sigma_{\mu i}^{(s)} - 2m_{\mathrm{del}}\cdot\rnd$, where $m_{\mathrm{del}}$ is the rate of deleterious mutations and $\rnd$ is a uniform random number in the range from 0 to 1. The parameter $m_{\mathrm{del}}$ is derived via the rate of beneficial mutations as described below.

The beneficial mutations are modeled as follows. With probability $f_{\mathrm{ben}} = 0.01$ at each step we add to $\sigma_{\mu i}^{(s)}$ the random quantity $B \cdot m_{\mathrm{ben}}$. The value $m_{\mathrm{ben}}$ is the stochastic effect of a single beneficial mutation on the harvesting effectiveness. It is generated randomly in the interval from 0 to 0.5 with exponential probability distribution $p(m_{\mathrm{ben}}) \propto \exp(-20m_{\mathrm{ben}})$. Parameter $B$ tunes the overall strength of genome corruption. For equilibrium (mutual vanishing of deleterious and beneficial mutation effects at large time scales), we require $B=1$, which leads to
\begin{equation}
  f_{\mathrm{ben}} \cdot \left< m_{\mathrm{ben}} \right> = m_{\mathrm{del}}.
\end{equation}
From the equation above one can obtain $m_{\mathrm{del}}=5.0\cdot10^{-5}$. The $B=0$ case corresponds to the total absence of beneficial mutations.

The total harvesting effectiveness of the species $\sum_i \sigma_{\mu i}$ is clipped between 0 and 1 to prevent the abundances from exceeding the resource influx and to reflect the fact that it is not possible to be specialized at harvesting all resources. Here, parallels with the expensive tissue hypothesis~\cite{navarrete2011energetics} can be drawn for illustration. The harvesting effectiveness $\sigma_{\mu i}$ is also restricted to take values below 0.

The developed model for beneficial and deleterious mutations follows generally the conclusions of Ref.~\cite{eyre2007distribution} and also lies in accordance with Fig.~6 from Chapter~7 of Ref.~\cite{sanford2012next}. To avoid extra parameters, the trait drift model does not include a dependence on population size, which is at least valid if the species have abundances on the same order of magnitude. This approximation is closely connected with the fixation probabilities of advantageous mutant genes~\cite{haldane1927mathematical,kimura1962probability,patwa2008fixation}; however, full correspondence cannot be drawn due to the finite time scales accounted in the present model.


\subsection{Speciation}
\label{subs_Spec}

To simulate speciation and forks in the evolutionary pathway, we divide the populations (species) when the size reaches a predefined threshold equal to $n=1$. The sub-threshold species can also split, but with a lower probability linear with abundance (a fork occurs if $n_{\mu} \geq \rnd$). This picture corresponds to the speciation described in Ref.~\cite{caldarelli1998modelling}. The species forks into two new species with the same harvesting effectiveness $\sigma_{\mu i}$, and the sum of the two new species abundance is equal to the initial abundance. After such division, we assume that complete reproductive isolation takes place between the species. However, the connection between the speciation rate and species abundance is still debatable due to the complex interplay of genetic drift, geographical ranges, bottlenecks, and mating behavior~\cite{drossel2001influence}. The concept of species in the present model corresponds most closely to the species belonging to the same genus or to a subpopulation of one species when mating is possible but is of low probability.

The species division procedure is performed at each fork (speciation) step, equal to $T=50$ unit steps. Examples of speciation time and the appearance of effective reproductive isolation on the order of several generations have been reported~\cite{armstrong2018precarious,belkina2018does}, and therefore the chosen value of $T$ is adequate. The fork time step $T=50$ is used in all simulations except for some where the effect of $T$ was investigated (see section~\ref{subs_decay}). The described mechanism corresponds better to prezygotic reproductive isolation. Unlike the Webworld model, the present model does not imply the distinction of ecological steps for determining the population dynamics and mutation/speciation steps. 


Plotting phylogeny trees in real life can be based on morphological comparisons of living species or obtained from the fossil record of extinct species and genome analysis. Here, for each species we introduce the genome and simulate it as an array of 100000 elements taking values 0 or 1. At each step, a random element in the array changes its value with a probability of $0.005$. A typical simulation consists of up to 100000 steps and thus approximately 500 elements are switched, which is sufficient for cluster analysis and plotting the phylogeny trees.

In the present model, the genome plays an auxiliary role in species relation analysis, and it contains no direct relation with the traits (i.e., harvesting effectiveness $\sigma_{\mu i}$). This situation is permissible because often the genome elements that are used to reconstruct the phylogeny (e.g., mitochondrial DNA~\cite{green2008complete,posth2017deeply,postillone2017mitochondrial,posth2016pleistocene} or Y chromosome~\cite{cruciani2011revised,wei2013calibrated}) do not have a direct connection with the traits that are crucial for the given period of evolution.




\section{Results}

\label{sec_res}

\subsection{Competitive exclusion principle}
\label{subs_competexcl}

The competitive exclusion principle is one of the most important ecological concepts, stating that species sharing the same resource cannot coexist at constant population values at a constant resource influx. Despite some investigations showing a violation the competitive exclusion principle in some specific conditions~\cite{PhysRevLett.118.028103}, we strictly require in the developed model reproduction of this principle.

First, it is instructive to focus exclusively on the part of the model responsible for population dynamics. We consider the case of a single resource with a constant influx in time and do not account for mutations or speciation. We compare the present model with that from Ref.~\cite{PhysRevLett.118.048103}. At the beginning, two species have slightly different harvesting effectiveness, 0.3 and $0.3\cdot(1+\varepsilon)$ ($\varepsilon\ll1$). For the present model, we fix the resource influx as $R_1=3.333$ and for the Tikhonov model as $R_1=1.0$, which gives the same limit value of abundance close to 1.0.

\begin{figure}
\includegraphics[width=0.49\textwidth]{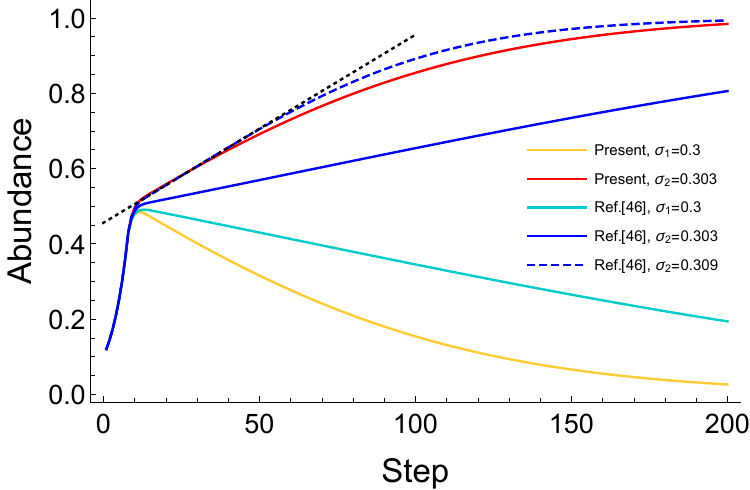}
\caption{\label{fig1} Competition for a shared resource by two species with close values of harvesting effectiveness. Yellow and red curves are for the time dependencies of the abundance of the \myf~and the \mys~species obtained by employing the present resource competition model. Their harvesting effectiveness values are $\sigma_1=0.3$ and $\sigma_2=0.303$, respectively. Cyan and blue curves are for the same for the model from Ref.~\cite{PhysRevLett.118.048103}. The blue dashed curve shows the abundance of the fitter species for a three times higher difference in harvesting effectiveness ($\sigma_1=0.3$ and $\sigma_2=0.309$) for the Ref.~\cite{PhysRevLett.118.048103} model. The black dashed line is a guide for the eye.}
\label{fig_compet0}
\end{figure}

Fig.~\ref{fig_compet0} shows the time dependencies of abundance in the described system. From the time steps 1 to 10, the total abundance does not reach the carrying capacity. As a result, the growth rate is not limited by resources, but only by the $f(x)$ giving the limit value of 20\% abundance growth per step. After reaching the carrying capacity, competition between the species starts. At the initial stages of competition, the rate of abundance divergence is proportional to the difference in harvesting effectiveness controlled by $\varepsilon$ for both models. However, in the present model, this rate is three times larger than that from the Ref.~\cite{PhysRevLett.118.048103} model. This is clearly seen if we increase $\varepsilon$ by three times in the latter model. One can conclude that the developed resource competition model brings stronger rivalry between the species. In many cases the two models behave in a similar manner, but in some situations the difference is significant and will be discussed.

The competitive exclusion principle is also obeyed in the full model accounting for population dynamics, mutations, and speciation. The corresponding simulations were designed as follows. Initially, two species exist. To distinguish between them and their descendants, their genomes were traced. The genome of the \myf~species was initially filled with zeros (``000...'' species) and the \mys~species genome was filled with ones (``111...'' species). A single resource with a constant in time influx of $R_1=2$ was taken. The initial harvesting effectiveness of the 000... and 111... species was $ \sigma_{11} = \sigma_{21} =  0.25$. These values of harvesting effectiveness and resource influx allow a total abundance of 0.5. The simulation was held for 5000 steps.

\begin{figure}
\includegraphics[width=0.45\textwidth]{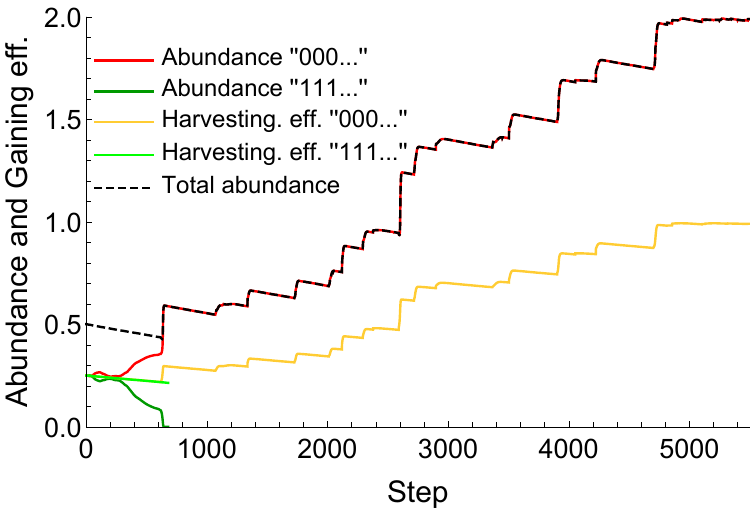}
\caption{Typical picture of interspecific competition illustrating the realization of the competitive exclusion principle in the present model. The total abundance of the 000... species is given in the red curve and that of the 111... species in the dark green curve. The corresponding mean harvesting effectiveness values $\sigma$ are depicted with yellow and light green. The black curve is for the total abundance in the ecosystem.}
\label{fig_compet}
\end{figure}

Fig.~\ref{fig_compet} shows typical dynamics of species abundance and harvesting effectiveness during the competition between species. At zero time, the 000... and 111... species had the same abundance, but due to stochasticity in weak beneficial mutations, the 000... species and its descendants were moderately winning the contest in the beginning. The total abundance in the ecosystem underwent small decreases over time due to the deleterious mutations. However, at approx. the 600$^{\rm th}$ step, a strong beneficial mutation in one of the species belonging to the group of 000... descendants occurred. As a consequence, this species acquired drastically better fitness and very rapidly spread in the ecosystem. All descendants of the 111... species became extinct.

The ecological time scales (dozens of steps in the present model) are smaller than those for evolution (thousands of steps). In real communities, a similar hierarchy of time scales takes place, see for example arguments in Ref.~\cite{drossel2001influence}. It is instructive to draw parallels with the real example of the process of \textit{Homo} genus evolution. The extinction of \textit{Homo  neandertalensis} took place at 30 kya \cite{higham2014timing} after the appearance of modern \textit{Homo sapiens} in Europe~\cite{trinkaus2003early,posth2017deeply} at 40--45 kya. The coexistence period is at least one order lower than the typical period of new species formation time for the \textit{Homo} genus (comparable with 100 kya). Another example is the rapid extinction of South American fauna after joining with North America due to continental drift~\cite{david2006great}. As a result, the chosen model parameters controlling the population dynamics and speciation ($f(x)$, beneficial and deleterious mutation rates, speciation time) allow the model to reproduce in a consistent manner this relation of ecological and evolutionary time scales.

\subsection{Resistance to the accumulation of deleterious mutations}
\label{subs_decay}

The most well-known manifestation of deleterious mutation accumulation is Muller's ratchet for asexual reproductive systems~\cite{loewe2006quantifying,CouceE9026}. Sexual reproduction is known as an efficient mitigation mechanism against the spread of deleterious mutations. Here, we show that the appearance of reproductive isolation and further competition between subpopulations also contribute to the resistance to the accumulation of deleterious mutations. Even in the situation when the impact of beneficial mutations is significantly weaker than that of deleterious mutations, total fitness in the ecosystem is shown to grow.

\begin{figure}
\includegraphics[width=0.45\textwidth]{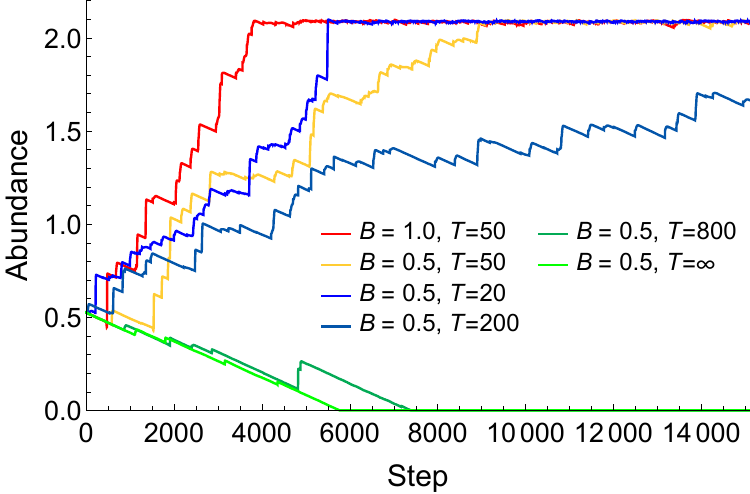}
\caption{\label{fig_Decay4curves} Time dependencies of total abundance in the ecosystem for various values of speciation time $T$ and parameter $B$ responsible for the strength of beneficial mutations. $B=0.5$ corresponds to a beneficial mutation impact two times lower than that for deleterious ones, and $B=1$ corresponds to the mutual vanishing of deleterious and beneficial mutation contributions.}
\end{figure}

Fig.~\ref{fig_Decay4curves} demonstrates the difference in the behavior of systems with various combinations of beneficial mutation rates (tuned by the parameter $B$) and speciation time $T$. For the simulation, we have used a single resource with an influx of $R_1=2.1$ and started with a single species with harvesting effectiveness $\sigma_{11} = 0.5$. When speciation is turned off ($T=\infty$), one in fact observes a bare drift of harvesting effectiveness due to the mutations, described in subsection \ref{subs_mutations}. In the case of a beneficial mutation impact two times lower than that for deleterious ones ($B=0.5$), the development of the ecosystem depends on speciation time $T$. At small $T$ the ecosystem does not degrade and reaches maximal adaptation (harvesting effectiveness close to 1) faster and abundance saturation. For a speciation time larger than approx. 200 for the given resource influx and beneficial mutations impact, the probability to become extinct is very high. 

From the obtained results, one can conclude that the acceleration of reproductive isolation emergence assists the mitigation of deleterious mutation effects. Qualitatively, the observed behavior can be explained by the appearance of competition between recently diverged populations/species. The population that stochastically gains more beneficial mutations and avoids deleterious ones will displace all other populations in the ecosystem. This conclusion is at least valid for the present model, where the time dynamics of fitness is considered to be independent of the population size. The fitness itself (presented here in the form of harvesting effectiveness $\sigma_{\mu i}$) should be contrasted with the mutation fixation rate known to be independent of the population size~\cite{haldane1927mathematical,kimura1962probability} and even of the reproductive isolation due to the population geographical structure~\cite{maruyama1974simple}. The time dynamics of fitness is governed by the interplay between the  emergence rate of mutations and the fixation probability and impact of the mutations. The obtained curves are consistent with those in Fig.~4 from Chapter~7 of Ref.~\cite{sanford2012next}. The usage of the resource competition model from Ref.~\cite{PhysRevLett.118.048103} results in similar behavior; the only difference is in the slower appearance of saturation due to weaker competition in resource harvesting.



\begin{figure}
\includegraphics[width=0.5\textwidth]{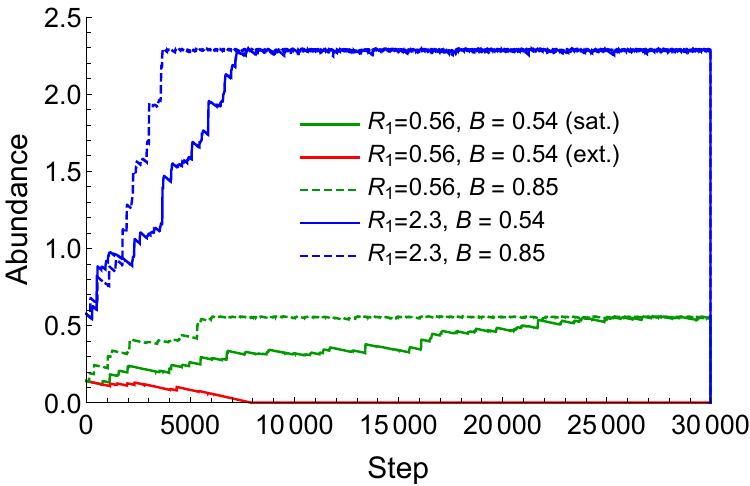}
\caption{\label{fig_speed5curves} Typical time dependencies of total abundance in the ecosystem for various values of resource influx $R_1$ and rate of beneficial mutations $B$. Adaptation is slower at lower beneficial mutation rates ($B=0.54$, solid curves) than at higher beneficial mutation rates ($B=0.85$, dashed curves). Adaptation is faster at higher resource influx ($R_1=2.3$, green) than at lower influx ($R_1=0.56$, black). At low values of $B$ and $R_1$, the probability of all species becoming extinct appears (red curve).}
\end{figure}

Fig.~\ref{fig_speed5curves} shows the typical time dependencies of the total abundance in the ecosystem at various values of resource influx and genome corruption strength due to deleterious mutations. One sees that at low levels of genome corruption (high $B$) and at high resource influx, evolution takes place faster, specifically the total abundance rapidly reaches the resource influx, which means that the mean harvesting effectiveness is saturated at the level of approx. 1.

\begin{figure}
\includegraphics[width=0.5\textwidth]{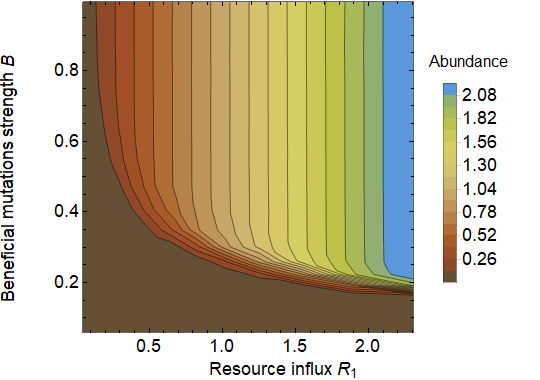}
\caption{\label{fig_phaseDecayResource} Phase diagram of extinction/survival: total abundance in the ecosystem as a function of beneficial mutations strength controlled by $B$ and total resource influx $R_1$. $B=0$ corresponds the case of no beneficial mutations and $B=1$ corresponds to mutual vanishing of deleterious and beneficial mutations. Brown area corresponds to the ecosystem with all species extinct.}
\end{figure}

Fig.~\ref{fig_phaseDecayResource} shows the abundance in the ecosystem as a function of resource influx $R_1$ and strength of beneficial mutations controlled by $B$ after 200000 steps. Averaging over 50 runs was performed for each point. Depending on the parameters, two possible scenarios take place. In the first, the protective mechanisms prevent the ecosystem from degradation, and the species evolve to the state of maximal adaptation: the harvesting effectiveness $\sigma$ saturates at the level of approx. 1. Thus the total abundance in the ecosystem reaches approximately the value of the resource influx $R$. The second scenario corresponds to an insufficient effect of beneficial mutations and consequently to global extinction. In Fig.~\ref{fig_phaseDecayResource} the brown area in the lower left corresponds to the parameters leading to such extinction.

\begin{figure}
\includegraphics[width=0.5\textwidth]{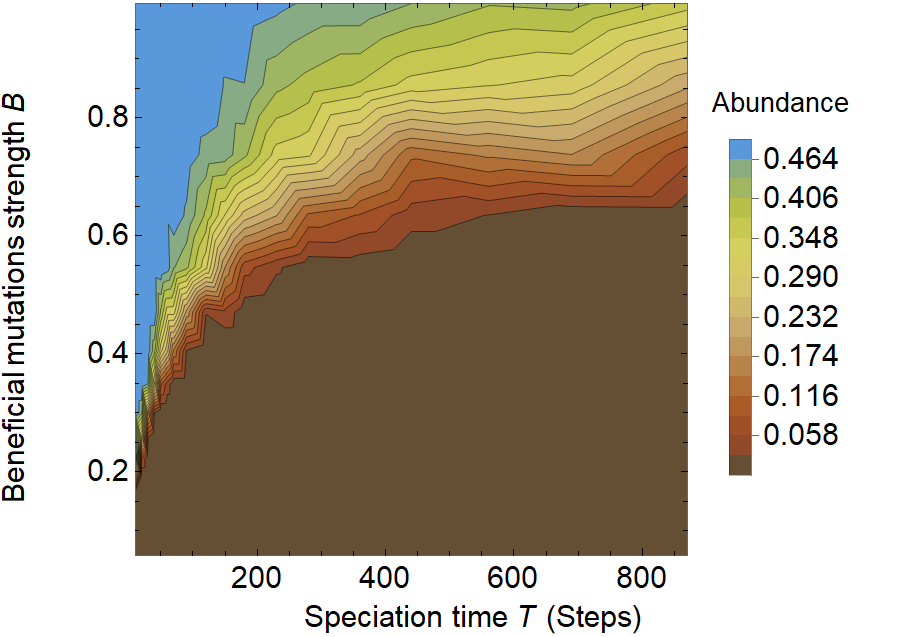}
\caption{\label{fig_phaseDecaySpecTime} Phase diagram of extinction/survival: total abundance in the ecosystem as a function of beneficial mutations impact $B$ and speciation (fork) time $T$. $B=0$ corresponds the case of no beneficial mutations and $B=1$ corresponds to the mutual vanishing of deleterious and beneficial mutations.}
\end{figure}

In the same way, the system can be explored in the phase space of the parameters of beneficial mutations strength $B$ and speciation (fork) time $T$ (cf. with Fig.~\ref{fig_Decay4curves}). The resource influx $R_1=0.5$ was taken. In Fig.~\ref{fig_phaseDecaySpecTime} one sees that damping the speciation leads to a decrease of the abundance in the ecosystem or even to the extinction of all species. On the contrary, acceleration of the emergence of reproductive isolation contributes to the mitigation of the negative impact of deleterious mutations.

\subsection{Filling empty ecological niches}
\label{subs_emptyniche}

In this subsection we demonstrate that our model behaves correctly in the situation of a new ecological niche appearing. We start from a single species of abundance 0.25 and harvesting effectiveness $\sigma_{11}=\sigma_{12}=0.25$. The influxes of the resources were initially $R_1=1.1$ and $R_2=0$. During the first 30000 steps, the influxes of resources remain constant. Then from 30000 to 50000 steps, $R_2$ grows linearly and reaches the value of $R_1$. Finally, until the end of the simulation (100000 steps), the influx of resources 1 and 2 are maintained at a level of 1.1.

\begin{figure}
\includegraphics[width=0.5\textwidth]{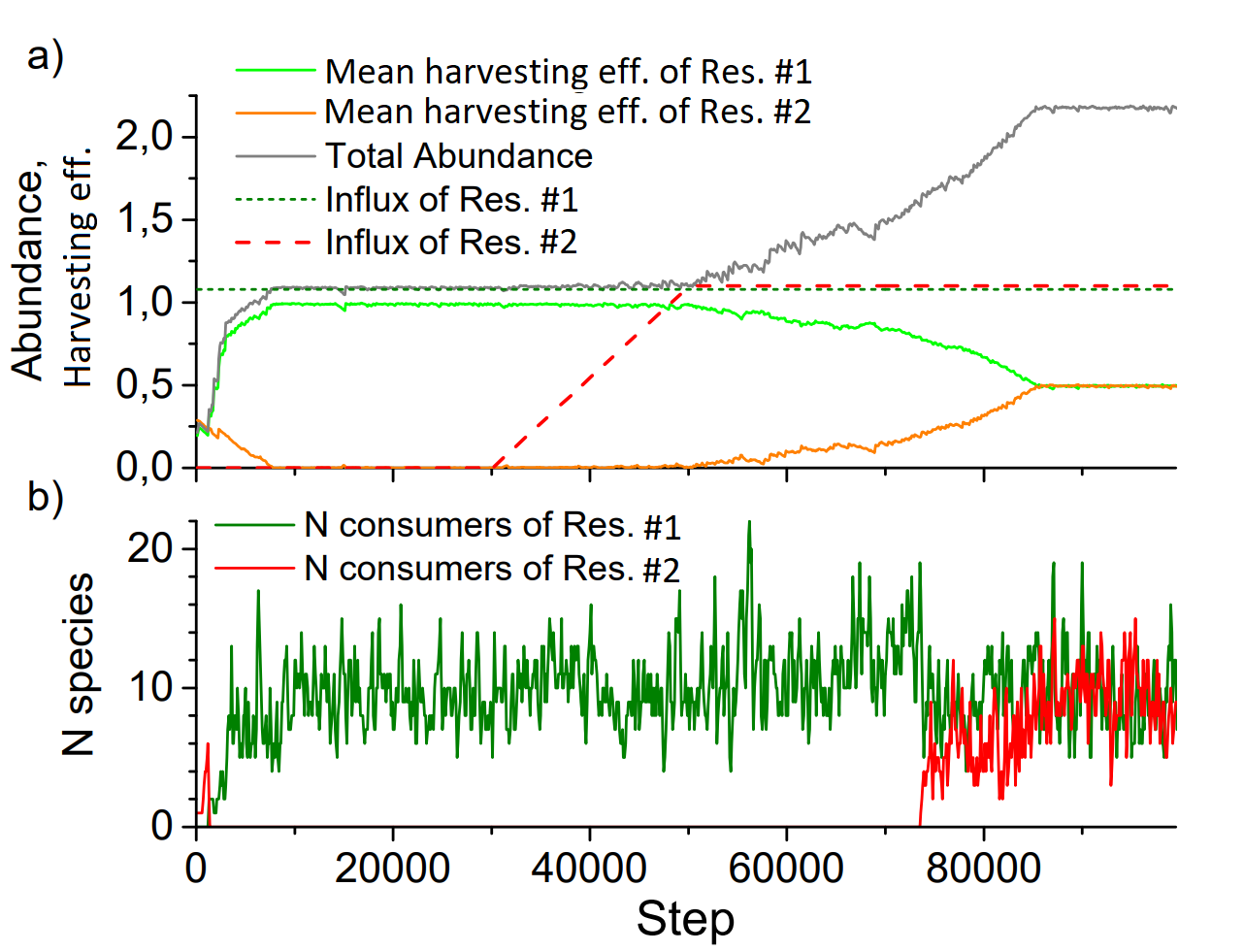}
\caption{\label{fig_newniche} Typical dynamics of harvesting effectiveness during the process of the appearance of a new ecological niche. (a) Growth from the 30000$^{\rm th}$ step of the \mys~ resource influx (red dashed curve) leads to a delayed increase of the corresponding mean harvesting effectiveness (orange curve). Simultaneously, the mean harvesting effectiveness of the \myf~ resource falls from 1 to 0.5 (light green curve). During the simulation, the influx of the \myf~ resource is constant (green dotted curve). Total abundance in the ecosystem grows (gray curve) due to the growth of total resource influx.(b) Number of species specialists for the \myf~and \mys~ resources plotted using green and red curves, respectively. }
\end{figure}

Fig.~\ref{fig_newniche} shows the process of filling the empty niche. During the first 10000 steps, rapid evolution and specialization of the species for the \myf~resource takes place. As a result the harvesting effectiveness $\sigma_{\mu1}$ saturates at the level of 1 and the total abundance also saturates at the level of $R_1=1.1$. At the end of the process of growth of the \mys~resource influx $R_2$, signatures of its harvesting become visible. By 75000 steps, the first specialists for the \mys~resource appear (the species with $\sigma_{\mu1}<\sigma_{\mu2}$). Finally by $T=85000$, specialists for the \mys~resource evolve and reach the maximal adaptation. The community is separated into two species groups specializing in the \myf~and \mys~resources, respectively. According to the molecular clock, the last common ancestor existed at approx. $T=50000$.

\subsection{Adaptive radiation at a large number of resource types}
\label{subs_specialization}

Here, the model is tested in complex conditions leading to a significantly richer behavior of the community. First, we consider an ecosystem with many resources ($N=5$), whose influx is constant in time. This condition should lead to specialization in one of the resources and an early divergence of species. The influx of the resources is $R_1=...=R_5=0.4$, providing a carrying capacity equal to 2.0. Simulation starts with a single species having a harvesting effectiveness of $\sigma_{1j}=0.05$ for $j=1,2,3,4,5$. The duration of the simulation was $100000$ steps.

\begin{figure*}
\includegraphics[width=0.59\textwidth]{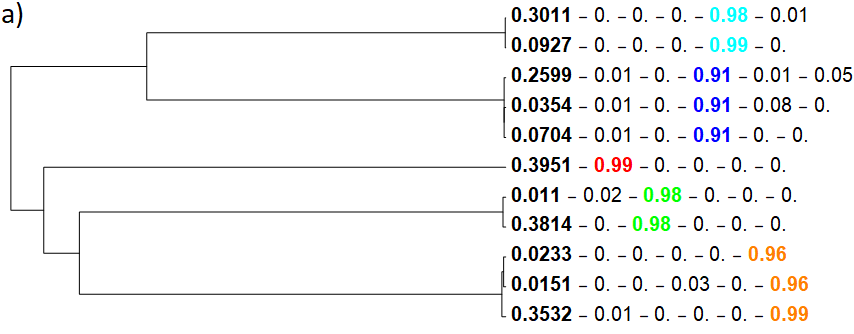}
\includegraphics[width=0.39\textwidth]{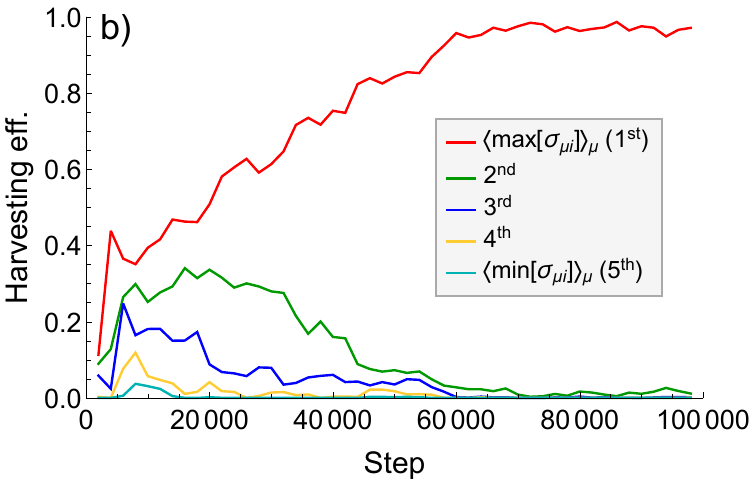}\\
\includegraphics[width=0.59\textwidth]{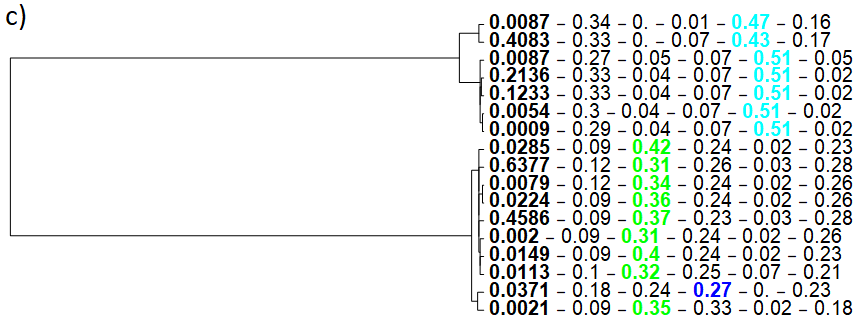}\includegraphics[width=0.39\textwidth]{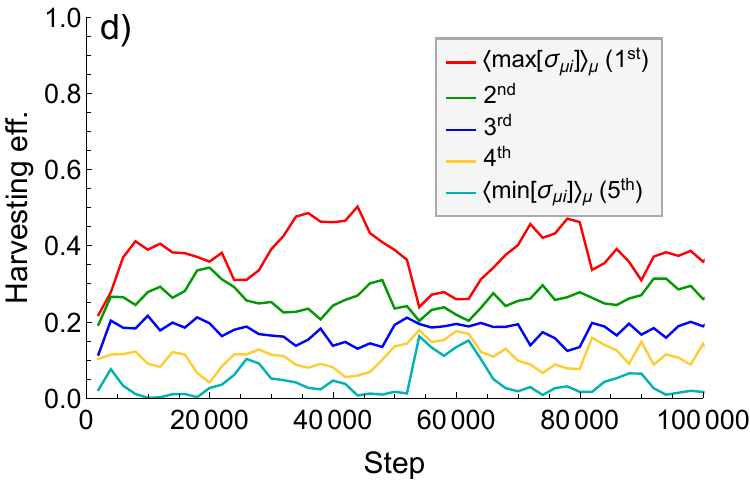}\\
\caption{\label{fig_dend_spec} Phylogeny trees and temporal dynamics of harvesting effectiveness for an ecosystem in a stable environment (constant in time and equal influx for all resources) for the developed resource competition model (panels a and b) and for the model from Ref.~\cite{PhysRevLett.118.048103} (panels c and d). The format of each leaf of the trees (panels a and c) provides the following information about the species abundance and harvesting effectiveness: $\mathbf{n_{\mu}} - \sigma_{\mu 1} - \sigma_{\mu 2} - \sigma_{\mu 3} - \sigma_{\mu 4} - \sigma_{\mu 5}$. The harvesting effectiveness for the preferable resource is highlighted with colors. The plots in (b) and (d) show the time dynamics of the harvesting effectiveness $\sigma_{\mu i}$ sorted for each species $\mu$  averaged over all species with weights given by species abundance. Red curves show the averaged maximal harvesting effectiveness. For the present model (b), the maximal harvesting effectiveness limit value is close to 1, whereas for the Ref.~\cite{PhysRevLett.118.048103} model (d) no such specialization takes place.}
\end{figure*}

Fig.~\ref{fig_dend_spec} shows a phylogeny tree of the species at the end of the simulation and the time dynamics of harvesting effectiveness for the present model and for the Tikhonov model. Tree leaves provide information about the harvesting effectiveness and abundance of the species. 

In the end of the simulation, the ecosystem consists of 5 groups of species (branches with similarly colored leaves in Fig.~\ref{fig_dend_spec}). Each of them contains several more or less successful species. The total abundance is close to the total resource influx, which means that the maximum possible adaptation was reached. This picture lies in agreement with the typical behavior of such ecological systems, where the number of consumers cannot overcome the number of available resources. Despite the fact that exceptions due to a specific complexity can occur in such situations~\cite{PhysRevLett.118.028103}, we judge that the equivalent number of large groups of species similar in specialization and the available resources lies in agreement with the competitive exclusion principle and no empty niche principle.

The divergence and specialization occurred very rapidly in this situation. According to the molecular clock technique, the last common ancestor existed at approximately the 3000$^{\rm th}$ step from the beginning (97000 steps previous). Meanwhile, within the group of close species specialized for the same resource, e.g., within the orange branch of the tree in Fig.~\ref{fig_dend_spec}a, divergence occurred 800 steps prior. Importantly, when the resource distribution from~\cite{PhysRevLett.118.048103} is used, the harvesting effectiveness demonstrates irregular patterns, and no specialization takes place during the simulation time. Thus, making specialization more favorable is an essential property of the developed resource competition model.

\begin{figure*}
\includegraphics[width=0.59\textwidth]{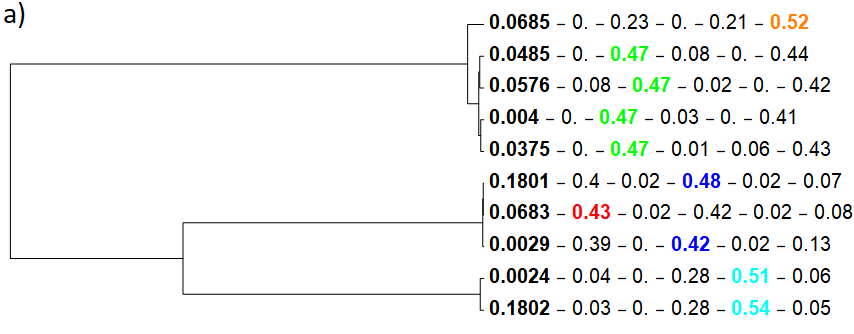}\includegraphics[width=0.39\textwidth]{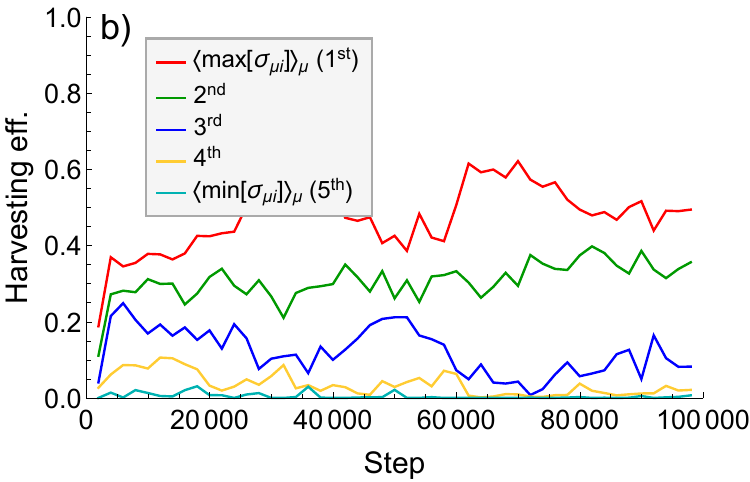}
\caption{\label{fig_dend_omn} (a) Phylogeny tree and (b) temporal dynamics of the harvesting effectiveness for an ecosystem in the situation of unstable resource influx.}
\end{figure*}

An alternative behavior can be expected in the case of an unstable environment~\cite{macarthur1967limiting} manifested in strongly fluctuating influx $R_i$ for all resources. To simulate this unstable environment, at each step a constant total resource influx of 2.0 was divided randomly into 5 parts to give the values of $R_{j}$. The other initial conditions were as in the previous case. One can expect no specialization for a single resource due to a high probability of extinction at the steps with a low influx of the preferable resource. Fig.~\ref{fig_dend_omn} shows a typical phylogeny plot for such situation along with the harvesting effectiveness time dynamics. All existing species become specialized in consuming two different resources. The total abundance was 0.65, which is lower than the total influx. The last common ancestor lived in the beginning of the simulation (approx. 5000$^{\rm th}$ step).

\section{Conclusion}

\label{sec_concl}

In the present study we proposed a model to describe the process of speciation and adaptive radiation governed by competition for resources and investigated its properties. The model accounts for resource competition and distribution, the effects of beneficial/deleterious mutations, and the appearance of reproductive isolation. The resource distribution model was developed based on the one from Ref.~\cite{PhysRevLett.118.048103}, but our model demonstrates a higher level of competition and more benefits from increasing harvesting effectiveness. Importantly, the present model allows examination of the impact of beneficial/deleterious mutations because the total abundance in the ecosystem is proportional not only to the resource influx but to the fitness (harvesting effectiveness $\sigma_{\mu i}$) of the species as well. As a result, extinction takes place not only due to competition with other species but also due to a decrease in fitness when the effect of deleterious mutations overcomes the effect of beneficial mutations.

Interspecific competition is due to sharing common resources without predator--prey relations. Thus, the developed approach describes species (or subpopulations) close in ecology that were recently subjected to reproductive isolation. The model reproduces in various conditions important ecological concepts: the competitive exclusion principle and the absence of vacant ecological niches.

The model highlights the transfer of the mechanism of selection from the level of individuals to the level of subpopulations, which helps to elucidate the widely disputed ways of preventing the effects of deleterious mutations from spreading in subpopulations and the process of increasing overall fitness during evolutionary. We show that a rapid appearance of reproductive isolation and a high resource influx can protect the ecosystem from deleterious mutations and are also favorable for overall adaptation due to the fixation of beneficial mutations. These conclusions coincide with the results obtained with the IBM approach~\cite{PhysRevE.69.051912}.

More complex systems were additionally studied, i.e., cases of a stable and unstable environment (resource influx) with a large number of accessible resource types. In the stable environment case, specialization was the preferable strategy for adaptation. The model reproduced lineage dynamics and reflected them in phylogeny trees, similar to what can be obtained from the fossil record. It was noteworthy that the usage of the resource competition model from Ref.~\cite{PhysRevLett.118.048103} could not lead to specialization in the stable environment. In the unstable environment case with a fluctuating resource influx, omnivory (i.e., absence of a single strongly preferred resource) was demonstrated as the optimal strategy.


Since the current model focuses on species at the same trophic level, competition for common resources takes place without considering predator--prey relationships. As a result, the simulated ecosystems did not exhibit the chaotic or cyclic behavior like that observed in the systems based on the Lotka--Volterra equations. The final ecosystem states in the present model were typically stable with the species optimally adapted to foraging the available resources.

Further development of the model can include connecting elements of the genome with harvesting effectiveness, adapting the developed resource competition model to the level of individuals in individual-based models, and incorporating consideration of the population geographical structure for studies of migrations, invasions, and allopatric speciation.


\section*{Acknowledgments}
The IBS Young Scientist Fellowship (IBS-R024-Y3-2022) is acknowledged for support. Thanks to D.D. Stupin and O. Bleu for useful remarks.


\bibliographystyle{ieeetr}
\bibliography{main}

\begin{thebibliography}{10}
\bibitem{misc1}
As an example, the resource influx can $R$ be measured in the units of joules per year per square kilometer and abundance in the units of individuals per square kilometer. The harvesting effectiveness $\sigma_{\mu i}$ is always dimensionless parameter in the range 0 to 1 showing the percentage of resource accessible by the species. In that case the resource requirement $\chi$ is to be measured in the units of joules per capita per year and the surplus $\Delta$ is to be normalized to $\chi$.
\end{thebibliography}

\end{document}